\documentclass[twocolumn,prl,superscriptaddress,showpacs,preprintnumbers,amsmath,amssymb]{revtex4}
\usepackage{graphicx}% Include figure files
\usepackage{dcolumn}% Align table columns on decimal point
\usepackage{bm}% bold math

\begin{document}

\preprint{}

\title{A direct experimental limit on neutron -- mirror neutron oscillations}

\author{G. Ban}
\affiliation{LPC Caen, ENSICAEN, Universit\'e de Caen, CNRS/IN2P3, Caen, France}
\author{K. Bodek}
\affiliation{Marian Smoluchowski Institute of Physics, Jagiellonian University, Cracow, Poland}
\author{M. Daum}
\affiliation{Paul Scherrer Institut (PSI), CH-5232 Villigen PSI, Switzerland}
\author{R. Henneck}
\affiliation{Paul Scherrer Institut (PSI), CH-5232 Villigen PSI, Switzerland}
\author{S. Heule}
\altaffiliation{Also at University of Z\"urich}
\affiliation{Paul Scherrer Institut (PSI), CH-5232 Villigen PSI, Switzerland}
\author{M. Kasprzak}
\altaffiliation{Also at SMI Vienna}
\affiliation{Paul Scherrer Institut (PSI), CH-5232 Villigen PSI, Switzerland}
\author{N. Khomutov}
\affiliation{JINR Dubna, Russia}
\author{K. Kirch} 
\altaffiliation{klaus.kirch@psi.ch}
\affiliation{Paul Scherrer Institut (PSI), CH-5232 Villigen PSI, Switzerland}
\author{S. Kistryn}
\affiliation{Marian Smoluchowski Institute of Physics, Jagiellonian University, Cracow, Poland}
\author{A. Knecht}
\altaffiliation{Also at University of Z\"urich}
\affiliation{Paul Scherrer Institut (PSI), CH-5232 Villigen PSI, Switzerland}
\author{P. Knowles}
\affiliation{University of Fribourg, Switzerland}
\author{M. Ku\'zniak}
\altaffiliation{Also at PSI Villigen}
\affiliation{Marian Smoluchowski Institute of Physics, Jagiellonian University, Cracow, Poland}
\author{T. Lefort}
\affiliation{LPC Caen, ENSICAEN, Universit\'e de Caen, CNRS/IN2P3, Caen, France}
\author{A. Mtchedlishvili}
\affiliation{Paul Scherrer Institut (PSI), CH-5232 Villigen PSI, Switzerland}
\author{O. Naviliat-Cuncic}
\affiliation{LPC Caen, ENSICAEN, Universit\'e de Caen, CNRS/IN2P3, Caen, France}
\author{C. Plonka}
\affiliation{Institut Laue Langevin, Grenoble, France}
\author{G. Qu\'em\'ener}
\affiliation{LPSC, Universit\'e Joseph Fourier Grenoble 1, CNRS/IN2P3, Institut National Polytechnique de Grenoble
53, avenue des Martyrs, 38026 Grenoble Cedex, France}
\author{M. Rebetez}
\affiliation{University of Fribourg, Switzerland}
\author{D. Rebreyend}
\altaffiliation{rebreyend@lpsc.in2p3.fr}
\affiliation{LPSC, Universit\'e Joseph Fourier Grenoble 1, CNRS/IN2P3, Institut National Polytechnique de Grenoble
53, avenue des Martyrs, 38026 Grenoble Cedex, France}
\author{S. Roccia}
\affiliation{LPSC, Universit\'e Joseph Fourier Grenoble 1, CNRS/IN2P3, Institut National Polytechnique de Grenoble
53, avenue des Martyrs, 38026 Grenoble Cedex, France}
\author{G. Rogel}
\altaffiliation{Also at LPC Caen}
\affiliation{Institut Laue Langevin, Grenoble, France}
\author{M. Tur}
\affiliation{LPSC, Universit\'e Joseph Fourier Grenoble 1, CNRS/IN2P3, Institut National Polytechnique de Grenoble
53, avenue des Martyrs, 38026 Grenoble Cedex, France}
\author{A. Weis}
\affiliation{University of Fribourg, Switzerland}
\author{J. Zejma}
\affiliation{Marian Smoluchowski Institute of Physics, Jagiellonian University, Cracow, Poland}
\author{G. Zsigmond}
\affiliation{Paul Scherrer Institut (PSI), CH-5232 Villigen PSI, Switzerland}

\date{\today}% It is always \today, today,

\begin{abstract}
In case a mirror world with a copy of our ordinary particle spectrum would exist,
the neutron n and its degenerate partner, the mirror neutron  ${\rm n'}$,
could potentially mix and undergo ${\rm nn'}$  oscillations. 
The interaction of an ordinary magnetic field with the ordinary neutron would
lift the degeneracy between the mirror partners, diminish the ${\rm n'}$-amplitude
in the n-wavefunction and, thus, suppress its observability.
We report an experimental comparison of
ultracold neutron storage in a trap with and without superimposed 
magnetic field. 
No influence of the magnetic field is found
and, 
assuming negligible mirror magnetic fields,
a limit on the oscillation time  
$\tau_{\rm nn'} > 103$\,s (95\% C.L.) is derived. 

\end{abstract}

%Use showkeys class option if keyword
\maketitle

The concept of a mirror world, as an attempt to restore global parity symmetry,
has attracted interest since the 1950's,
started by the famous paper of Lee and Yang~\cite{Lee56}
and significantly expanded in the work of Kobzarev, Okun, and Pomeranchuk~\cite{Kob66}.
The mirror matter idea was first applied to the Standard Model of particle physics in~\cite{Foo91}.
More recent overviews can be found in~\cite{Ber04,Oku06}.
The mirror world could hold a 
copy of the particle spectrum of our ordinary world. 
Matter and mirror matter would interact via gravity
and present a viable explanation to the dark matter 
problem~\cite{Bli82,Foo04a,Ber05,Foo04b,Foo06}.
Besides gravity, other (new) interactions could show up in minute mixings 
of neutral matter particles --- such as neutrinos, pions, kaons, or 
positronium (see~\cite{Bad07} for $e^+e^-$) --- 
and their degenerate mirror partners making oscillations between them possible. 
Recently it was pointed out~\cite{Ber06a} that no direct experimental limits exist on the
oscillation time  $\tau_{\rm nn'}$~\cite{tau_def}
between ordinary matter neutrons (n) and the speculative mirror neutrons (${\rm n'}$).
An indirect limit of the order $\tau_{\rm nn'}\geq1$\,s has been derived in~\cite{Ber06a}
based on the search for neutron -- antineutron (n$\bar{\rm n}$) oscillations~\cite{Bal94}.
Fast n${\rm n'}$ oscillations with $\tau_{\rm nn'}\sim1$\,s, or at least much shorter than the neutron 
$\beta$-decay lifetime,
could explain~\cite{Ber06a,Ber06b}
the origin of ultra-high energy cosmic rays above the Greisen-Zatsepin-Kuzmin (GZK) cutoff~\cite{Gre66,Zat66}.
The viability of models and implications have been further discussed in~\cite{Moh05}.

Possible approaches to n${\rm n'}$ oscillation experiments with sensitivities 
of several hundred seconds have been discussed in~\cite{Pok06}.
One approach is to search for n${\rm n'}$ oscillations by comparing the storage of ultracold
neutrons (UCN) in vacuum in a trap in the presence and the absence, 
respectively, of a magnetic field. 
The essential idea is that the
neutron and mirror neutron states would be degenerate in the absence of a magnetic field
and n${\rm n'}$ transitions could occur. 
(The absence of mirror magnetic fields at the location of the experiment
is assumed throughout this paper~\cite{footnote_mirrmag}.)
The interaction of the neutron with a 
magnetic field would lift 
the degeneracy and suppress the transition into a mirror neutron
which, of course, does not interact with the ordinary magnetic field,
nor with the trap via the ordinary strong interaction.
Thus, the oscillation into mirror neutrons adds a loss channel
for ultracold neutron storage. 
If n${\rm n'}$ transitions occurred,
the storage time constant for ultracold neutrons in a trap with
magnetic field would be longer than without magnetic field.
One should note that this disappearance method only measures neutron
loss as a function of applied magnetic field.  A~signal will not prove
the oscillation into mirror neutrons, only that some magnetic field
dependent loss channel exists.  By assuming that the n${\rm n'}$
oscillation is responsible for the loss, limits can be set on
$\tau_{\rm nn'}$.
One can imagine
other exotic disappearance channels
for the neutron, among which only the antineutron channel
is tightly constrained~\cite{Bal94}.

The formulation of the n${\rm n'}$ oscillation is analogous to the evolution of other
simple two state systems 
such as spin $\frac{1}{2}$, K$^0\bar{\rm K}^0$, or  n$\bar{\rm n}$ mixing (see, e.g.,~\cite{Coh77,Moh80}). 
The energy difference between neutron and mirror neutron states due to 
magnetic field interaction with the neutron magnetic moment $\mu$ is $\mu B$.
For convenience we adopt the notation of~\cite{Pok06,Ign90}
and define a characteristic frequency $\omega \equiv \frac{\mu B}{2\,\hbar}$ which corresponds to
half the energy splitting.
The probability $p$ for an UCN to be found as a mirror neutron after
a time $t$ can then be written as
\begin{equation}
p(t) = \frac{{\rm sin}^2(\sqrt{1+(\omega \tau_{\rm nn'})^2}\times t/\tau_{\rm nn'})}{1+(\omega \tau_{\rm nn'})^2}\,.
\label{np_eqn}
\end{equation}
The time $t$ is limited by the free flight time $t_f$ between two wall collisions.
The wall collision frequency is determined by $\frac{1}{t_f}$. 
The effective transition rate of UCN into mirror neutrons is then given by
\begin{equation}
R = \frac{1}{t_f} \int\limits_0^{t_f} \frac{dp}{dt} \, dt
  = \frac{1}{t_f} \, p(t_f)\,.
\label{Rp_eqn}
\end{equation}
For a real system the factors on the right hand side of Eq.~(\ref{Rp_eqn}) 
must be properly averaged over the distribution of flight times between collisions
during the storage time $t_s$:
\begin{equation}
\label{Rs_eqn}
R_s =\frac{1}{\langle t_f \rangle_{t_s}}  \,  \langle p(t_f)\rangle_{t_s}\,.
\end{equation}
In experiments, one searches for a weak coupling, thus long $\tau_{\rm nn'}$:
so in practice $\omega \tau_{\rm nn'} \gg 1$ in Eq.~(\ref{np_eqn}).
Two limits are considered for Eq.~(\ref{np_eqn}):
In the first case (``${\uparrow\!\downarrow}$''), 
$\omega_{\uparrow\!\downarrow} t_f \gg 1$ (large B-field), many oscillations take place and 
the ${\rm sin}^2(..)$ term of Eq.~(\ref{np_eqn}) is averaged to $\frac{1}{2}$
because $t_f$ varies along UCN trajectories:
\begin{equation}
\label{nRA_eqn}
R_{s,{\uparrow\!\downarrow}} 
= \frac{1}{\langle t_f \rangle_{t_s}}  \, 
\frac{1}{2(\omega_{\uparrow\!\downarrow} \,\tau_{\rm nn'})^2}\,.
\end{equation}
In the second case  (``$0$''), 
$\omega_0 t_f \ll 1$ (small B-field), the ${\rm n'}$ component grows quadratically 
in time during the free flight:
\begin{equation}
\label{nRB_eqn}
R_{s,0}  = \frac{1}{\langle t_f \rangle_{t_s}}  \, 
\frac{\langle t_f^2\rangle_{t_s}}{\tau^2_{\rm nn'}}\,.
\end{equation}

Also regular losses of UCN must be considered, such as 
absorption and upscattering (during wall interactions or 
in collisions with rest gas), trap leakage and $\beta$ decay.
All these loss mechanisms contribute to the UCN 
storage time constant $\tau_{\rm store}$ of the system; $\lambda_{\rm store}=1/\tau_{\rm store}$
is the corresponding loss rate. Generally, the loss rate depends on UCN energy and
for a spectrum of stored UCN the decay curve is a sum of exponentials. 
The total effect can be modeled by the relative populations  $c_i$ of different
velocity classes, each with its own storage loss rate $\lambda^{(i)}$.
After storing an initial number $n(t\!\!=\!\!0)$ of UCN
for some time $t_{s}$ in a given magnetic field one will
detect the number of surviving UCN
\begin{equation}
\label{sumexp_eqn}
n(t_s)=n(t\!=\!0) \times \sum_i{ c_i\,{\rm exp}[-(\lambda^{(i)}_{\rm store}+R_s)\,t_s]}\,.
\end{equation}
with the simple normalization condition $\sum_i c_i = 1$.
For measurements in the limits ${\uparrow\!\downarrow}$ and $0$ (only the magnetic field is changed),
the ratio of detected UCN becomes independent of all the regular UCN
loss mechanisms
\begin{equation}
\label{NBA}
N_{0/{\uparrow\!\downarrow}}\equiv 
\frac{n_{\rm 0}(t_s)}{n_{\rm {\uparrow\!\downarrow}}(t_s)}
= {\rm exp}[(R_{s,{\uparrow\!\downarrow}}-R_{s,0})\,t_s]\,.
\end{equation}
In the absence of other effects, 
neutron -- mirror neutron oscillations lead to $N_{0/{\uparrow\!\downarrow}}<1$.

We have performed UCN storage experiments at the Institut Laue-Langevin using the
experimental setup of the neutron EDM experiment~\cite{Har99,Bak06}.
A typical measurement cycle consists of (i) a filling time of 40\,s with the beam switch connecting
the storage chamber to the ILL PF2 EDM beam line~\cite{Ste86} 
allowing {\it unpolarized} UCN to enter the storage volume, (ii) 
different storage times $t_s$ when the UCN isolation shutter to the
storage chamber was closed
and (iii) 40\,s counting time with the UCN shutter open and the beam switch connecting the storage chamber
to the $^3$He filled UCN detector~\cite{Strelkov}. 
The pressure inside the storage chamber was always better than
10$^{-3}$\,mbar in order to make sure that the ${\rm nn'}$
degeneracy is not lifted by the interaction of UCN with the
rest gas.

The UCN storage chamber has a volume $V\sim 21$~l and a surface area
of $A\sim5400$\,cm$^2$.  The limit for stored UCN~velocity is 4.1\,m/s.
From kinetic gas theory, the mean free path of UCN between wall
collisions is $\frac{4\,V}{A}\approx0.16$\,m, the mean velocity is
about 3\,m/s~\cite{Pen04} and, thus, $\langle t_f \rangle
\approx0.053$\,s.  One obtains $\omega \cdot \langle t_f\rangle \sim
1$ at a magnetic field of $B\sim0.42\,\mu$T; the limiting cases are
obtained for magnetic fields of more than a few $\mu$T
(${\uparrow\!\downarrow}$) and of less than a hundred nT (0),
respectively.

Different magnetic field configurations were used:
up ($B_\uparrow$), off ($B_0=0$) and down ($B_\downarrow$). The strength of the magnetic field was adjusted
by the current through the main magnetic field coil. The relevant measurements 
were taken at $|B_{\uparrow\!\downarrow}|\approx6$\,$\mu$T (100\,mA). 
The magnetic field ($B_z$, along the main magnetic field direction)
as a function of the applied current was measured using the 
Hg cohabiting magnetometer~\cite{Gre98} for fields between 0.3\,$\mu$T and 7\,$\mu$T.
For lower fields the Hg magnetometer could not be used.
The Hg data set shows a perfectly linear dependence of the field on applied current and
results in $|B_{z,0}|=2.3\pm2.6$\,nT when extrapolated to zero current. 
This value for the magnetic field along the main field axis indicates
a residual absolute B-field below 13\,nT because of the absence
of a preferred spatial direction.
The zero field $B_0$ for the actual measurements was set by switching off the
coil current and demagnetizing the four-layer mu-metal shield surrounding the storage chamber.
We also used 3-axis fluxgate sensors 
directly above the storage chamber in order to verify that the residual B-field was
sufficiently small for the purpose of our experiment. 
The B-field configurations of the experiment  were very well reproducible, and in particular $B_0$
within less than 1\,nT.
The direct limit on $|B_0|$ obtained from the fluxgates is,
however, somewhat weaker: it was found that the connectors of the commercial devices
are slightly magnetic, leading to offset fields at the location of the sensor on the order
of 25\,nT. 
Although the residual field is probably on the level of a few nT, we give a conservative
limit of  $|B_0|<50$\,nT, which is sufficient for our purpose here, i.e., for the limiting case ``0''.

Most of the measurements were performed repeating the sequence (a) 
$(B_0,B_\uparrow,B_\downarrow,B_0,B_0,B_\downarrow,B_0,B_\uparrow)$
with field changes typically every 1.5\,h during day time. 
The demagnetization procedure 
before $B_0$ measurements took about half an hour. 
For a given B-field configuration 16 UCN cycles were measured: 4 for each
storage time of $t_s=100$\,s, 50\,s, 175\,s, and again 100\,s. 
Night runs were taken for longer periods at one B-field configuration
with $t_s=100$\,s. They were
used to check on the long term stability of the system. 
It was found that drifts of the
count rates were slow and on a level below 0.3\% over
several hours. This agrees in magnitude with changes in the
reactor power, but a direct correlation could not be established.
The count rate drifts were sufficiently slow to be averaged out
in the day runs with frequent changes of the B-field configuration.
For $t_s=50$\,s some data was taken using another sequence (b),
$(B_\uparrow,B_\downarrow,B_0)$ while checking on an unexpected 
count ratio $N_{0/{\uparrow\!\downarrow}}$ (see below).

The time constant for UCN to leave the storage chamber with the shutter open
was measured to be $\tau_{op}=11.4\pm0.6$\,s. 
Mirror neutrons can leave the system also during filling
and counting. The relevant average times $t_s^*$ 
in our storage system are thus longer than the times 
between closing and opening the UCN shutter. 
One can replace $t_s$ in Eq.~(\ref{NBA}) by $t_s^*$ and,
because $R_s$ changes only rather weakly with $t_s$
and $\tau_{op}$ is much smaller than $t_s$, one finds  
to very good approximation  $t_s^*=t_s + 2 \, \tau_{op}$.
We assign a conservative systematic error of $\pm3$\,s to $t_s^*$.

\begin{table}
\begin{center}
\vspace{1cm}
\begin{tabular}{c||c|c|c|c}

$t_s\,[\,{\rm s}\,]$         &50 (a)             &50 (b)     &100 (a)              &175 (a)\\ 
$t_s^*\,[\,{\rm s}\,]$         &$73\pm3$ (a)    &$73\pm3$ (b)     &$123\pm3$ (a)    &$198\pm3$ (a)\\ \hline
%&&&&\\
$n(B_0)$        &$44317\pm40$              &$44363\pm53$            &$28635\pm21$         &$17015\pm22$\\
%&&&&\\
$n(B_\uparrow)$   &$44197\pm53$              &$44443\pm53$            &$28671\pm30$         &$17047\pm31$\\
$n(B_\downarrow)$ &$44128\pm53$            &$44316\pm46$            &$28596\pm30$         &$16974\pm31$\\
%&&&&\\
$n(B_{\uparrow\!\downarrow})$ &$44163\pm38$ &$44371\pm35$            &$28633\pm22$         &$17011\pm22$\\ 
%&&&&\\
\hline
%&&&&\\
$N_{0/{\uparrow\!\downarrow}}$ &$1.0035(13)$  &$0.9998(15)$            &$1.0001(11)$        &$1.0002(18)$\\ 
&\multicolumn{2}{c|}{1.0019(10)}&&\\

\end{tabular}
\caption{\label{result_tab}
Measured total UCN counts $n$ normalized per cycle for the day sequences ((a), (b), see text)
at different storage times $t_s^*$ (with systematic error)
and magnetic field configurations.
$n(B_{\uparrow\!\downarrow})$ is the weighted average of $n(B_\uparrow)$ and $n(B_\downarrow)$, 
and $N_{0/{\uparrow\!\downarrow}}=n(B_0)/n(B_{\uparrow\!\downarrow})$. 
}
\end{center}
\end{table}

For data analysis, each B-field configuration was first treated separately.
The distributions of single-cycle counts $n$ were found to be consistent with Gaussians
with standard deviations $\sigma \approx \sqrt{n}$ with no additional systematics.
The counts per cycle for each configuration were thus statistically averaged,  
see Table~\ref{result_tab}. The averaged $n$ data
show a presently unexplained tendency to a
linear dependence on the magnetic field.  The effect we wish to limit
depends on $|B|^2$ (via $\omega^2$), so the direct average of the $+6$
and $-6$~$\mu$T measurement values cancels the linear systematic
effect, leaving only the possible oscillation effect and any remaining
quadratic systematic contributions.
The averaged $n$ data is then used to calculate the count ratios
$N_{\rm 0/{\uparrow\!\downarrow}}$ according to Eq.~(\ref{NBA}). 
For 50\,s storage time and sequence (a), the count ratio $N_{0/{\uparrow\!\downarrow}}$
is larger than 1 by 2.7 standard deviations, which led us to remeasure at this storage time
(using sequence (b)) and corroborate that this deviation was a statistical fluctuation.
Both ratios $N_{\rm 0/{\uparrow\!\downarrow}}$(50\,s) are given in the table along with the
obtained average.
The individual results for $N_{\rm 0/{\uparrow\!\downarrow}}$ show no
signal within their respective sensitivities and, as they are independent,
can be used in a combined analysis.
Following Eq.~(\ref{NBA}), we write
$N_{0/{\uparrow\!\downarrow}}= {\rm exp}[a \, t_s]$ with a fit parameter $a$. 
The fit gives $a=(5.38\pm5.78)\times 10^{-6}~\mathrm{s}^{-1}$.
We use this value at the limit of the experimentally measured
range ($t_s^*\!=\!198$\,s) to set the constraint on the neutrons which
may have been lost, yielding
\begin{equation}
N_{0/{\uparrow\!\downarrow}}(t_s^*\!=\!198\,{\rm s})=1.00106 \pm 0.00114\,.
\end{equation}
Results with $N_{0/{\uparrow\!\downarrow}}>1$ are unphysical for ${\rm nn'}$ oscillations.
In order to derive a limit on
$N_{0/{\uparrow\!\downarrow}}$ we adopt the Bayesian approach
described by the Particle Data Group (page~305 of Ref.~\cite{PDG06};
probability distribution set to zero for $N_{0/{\uparrow\!\downarrow}}>1$)
and obtain
\begin{equation}
\label{Nlim}
N_{0/{\uparrow\!\downarrow}}(t_s^*=198\,{\rm s})>0.99835\,{\rm s \,(95\% C.L.)}. 
\end{equation}

In order to derive the limit on $\tau_{\rm nn'}$,
the flight time distribution averages
$\langle t_f \rangle_{t_s}$ and  $\langle t_f^2 \rangle_{t_s}$ are needed
as additional input.
A better determination than the one from the kinetic gas theory argument given above
was obtained by Monte Carlo calculations using Geant4UCN~\cite{G4UCN}.
The parameters of the  simulation (mainly material properties, such as 
Fermi pseudo-potential, loss probability per wall reflection, and 
fraction of diffuse to specular reflection)  
have been tuned to reproduce measurements of the UCN beam energy spectrum and
filling, storage, and emptying time curves of the apparatus.
Excellent agreement with the observables is obtained, which justifies the
extraction of the required flight time distributions. The averages are
given in Table~\ref{MC_tab}. The assigned systematic uncertainties 
were derived by varying the material parameters of the simulation, the
largest influence coming from the loss probabilities per wall collision.

\begin{figure}
\includegraphics[width=1.\linewidth, angle=0]{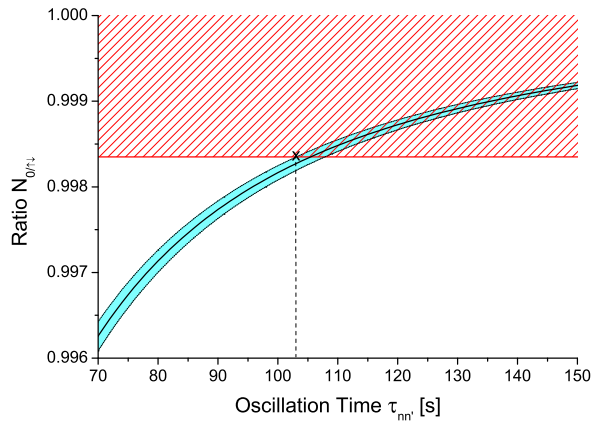}
\caption{\label{Figure_1.eps}
(color online) The count ratio $N_{0/{\uparrow\!\downarrow}}$ as a function of 
the oscillation time (see Eq.~(\ref{NBA})). The dashed region indicates the allowed
region of Eq.~(\ref{Nlim}). 
The error band of the curve displays the systematic uncertainties, see text. The X indicates
the point at which the limit on $\tau_{\rm nn'}$ is evaluated.
}
\end{figure}

The limit on the oscillation time is obtained using the limit on lost
neutrons, Eq.~(\ref{Nlim}),
and the average free flight time values (at $t_s=175$\,s, from Table~\ref{MC_tab})
in Eq.~(\ref{NBA}) and solving for $\tau_{\rm nn'}$.
The systematic uncertainties are taken
into account for the limit on $\tau_{\rm nn'}$
by adding (or subtracting) them simultaneously in order for them to weaken the limit,
i.e., $\langle t_f \rangle_{t_s}=0.0548$\,s, $\langle t_f^2 \rangle_{t_s}=0.00515$\,s$^2$ and $t_s^*=195$\,s.
%We obtain the final result
With a reminder of the assumptions (negligible mirror
magnetic field~\cite{footnote_mirrmag}, no conventional strong or electro--magnetic 
interactions of the mirror neutron ${\rm n'}$, and degeneracy of ${\rm n}$ and 
${\rm n'}$ in the gravitational field) we obtain the final result
\begin{equation}
\tau_{\rm nn'}> 103\,{\rm s \,(95\% C.L.)}\,. 
\end{equation}
Figure~\ref{Figure_1.eps} 
displays the dependence of $N_{0/{\uparrow\!\downarrow}}$ on $\tau_{\rm nn'}$ (see Eq.~(\ref{NBA}))
with the band indicating the influence of the systematic uncertainties.
The 95\% confidence limit of Eq.~(\ref{Nlim}) is shown and the cross marks the point which determines
the limit on the oscillation time. 

The result impacts the role ${\rm nn'}$ oscillations can play in the
transport of ultra-high energy cosmic rays over large distances~\cite{Ber06a,Ber06b}, although, 
it may not completely rule out the ${\rm nn'}$-explanation for events above the GZK cutoff.

\begin{table}
\begin{center}
\vspace{1cm}
\begin{tabular}{c||c|c|c}
$t_s\,[\,{\rm s}\,]$         &50  &100            &175\\ 
\hline
%&&&\\
$\langle t_f \rangle_{t_s}\,[\,{\rm s}\,]$  &0.0498(5) &0.0515(5) &0.0543(5)\\
%&&&\\
$\langle t_f^2\rangle_{t_s}\,[\,{\rm s^2}\,]$ 
&0.00420(8) &0.00450(9) &0.00505(10) \\

\end{tabular}
\caption{\label{MC_tab}
Results for $\langle t_f \rangle_{t_s}$ and $\langle t_f^2 \rangle_{t_s}$
using Monte Carlo distributions of flight times between wall collisions.
Variation of parameters in the simulation is used to assign systematic uncertainties. 
}
\end{center}
\end{table}

This work was performed at the 
Institute Laue-Langevin, Grenoble, France.
We are grateful to our colleagues of the RAL--Sussex--ILL collaboration~\cite{Bak06}
for the loan of experimental equipment and many fruitful discussions.
Discussions with V. Nezvishevsky led us to start the mirror neutron investigation.
We acknowledge the great support of technical services
throughout the collaboration.
This work is supported by grants from the Polish Ministry of Science
and Higher Education, contract No.~336/P03/2005/28,  and the Swiss
National Science Foundation \#200020--111958.


\begin{thebibliography}{00}
\bibitem{tau_def}
       In the absence of any external interaction lifting the
       ${\rm nn'}$ degeneracy,
       $\frac{\pi}{2}\,\tau_{\rm nn'}$ is the time
       after which one finds with 100\% probability a mirror neutron from
       an initial neutron.
\end{thebibliography}
\end{document}